\documentclass[pre,longbibliography]{revtex4-1}

\ifx\pdfoutput\undefined
\usepackage{graphicx}
\else
\usepackage[pdftex]{graphicx}
\usepackage{epstopdf}
\fi

\usepackage[cp850]{inputenc}
\usepackage{amsmath,amssymb,color}
\usepackage{graphicx}
\usepackage{latexsym}
\usepackage{psfrag}
\usepackage{color}
\usepackage{wasysym}
\usepackage{epsfig}

\providecommand\bnabla{\boldsymbol{\nabla}}
\providecommand\btau{\boldsymbol{\tau}}

\newcommand*{\vcenteredhbox}[1]{\begingroup
\setbox0=\hbox{#1}\parbox{\wd0}{\box0}\endgroup}

\newcommand{\textin}[1]{\mbox{\scriptsize{#1}}}

\begin{document}

\title{Influence of the surface viscous stress on the pinch-off of free surfaces loaded with nearly-inviscid surfactants}

\author{A. Ponce-Torres$^1$, M. Rubio$^1$, M. A. Herrada$^2$, J. Eggers$^3$ and J. M. Montanero$^1$\\
$^1$ Depto.\ de Ingenier\'{\i}a Mec\'anica, Energ\'etica y de los Materiales and\\
Instituto de Computaci\'on Cient\'{\i}fica Avanzada (ICCAEx),\\ Universidad de Extremadura, E-06006 Badajoz, Spain\\
$^2$ Depto.\ de Mec\'anica de Fluidos e Ingenier\'{\i}a Aeroespacial,\\
Universidad de Sevilla, E-41092 Sevilla, Spain\\
$^3$ School of Mathematics, University of Bristol, Fry Building, Bristol BS8 1UG, UK}

\begin{abstract}
We analyze both theoretically and experimentally the breakup of a pendant water droplet loaded with Sodium Dodecyl Sulfate (SDS). The free surface minimum radius measured in the experiments is compared with that obtained from a numerical solution of the full Navier-Stokes equations for different values of the shear and dilatational surface viscosities. This comparison shows the small but measurable effect of the surface viscous stresses on the system dynamics for sufficiently small spatiotemporal distances from the breakup point, and allows to establish upper bounds for the values of the shear and dilatational viscosities. We study numerically the distribution of Marangoni and viscous stresses over the free surface as a function of the time to the pinching, and describe how surface viscous stresses grow in the pinching region as the free surface approaches its breakup. When Marangoni and surface viscosity stresses are taken into account, the surfactant is not swept away from the thread neck in the time interval analyzed. Surface viscous stresses eventually balance the driving capillary pressure in that region for small enough values of the time to pinching. Based on this result, we propose a scaling law to account for the effect of the surface viscosities on the last stage of the temporal evolution of the neck radius.
\end{abstract}

\maketitle

\section{Introduction}
\label{sec1}

Soluble surfactants play a fundamental role in many microfluidic applications \citep{A16}. For instance, it is well-known that surfactants can stabilize both foams and emulsions due to Marangoni convection effects \citep{EBW91,CT04,DL08}. The surface viscosity of surfactant monolayers is also believed to play a significant role in such stabilization. In fact, the drainage time during the coalescence of two bubbles/droplets can considerably increase due to the monolayer viscosity \citep{OJ19}. However, there are serious doubts about whether small-molecule surfactants commonly used in microfluidic applications exhibit measurable surface viscosities. For instance, \citet{ZNMLDMTS14} reported that the surface shear viscosity of Sodium Dodecyl Sulfate (SDS) was below the sensitivity limit of their experimental technique ($\sim 10^{-8}$ Pa\,s\,m). This raises doubts about the role played by surface shear rheology in the stability of foams and emulsions treated with soluble surfactants.

The disparity among the reported values of shear and dilatational viscosities of both soluble and insoluble surfactants reflects the complexity of measuring such properties. The lack of precise information about these values, as well as the mathematical complexity of the calculation of the surface viscous stresses, has motivated that most of the experimental and theoretical works in microfluidics do not take into account those stresses. However, one can reasonably expect surface viscosity to considerably affect the dynamics of interfaces for sufficiently small spatiotemporal scales even for nearly-inviscid surfactants \citep{PMHVV17}. A paradigmatic example of this is the pinch-off of an interface covered with surfactant \citep{PMHVV17}, where both the surface-to-volume ratio and surface velocity can diverge for times and distances sufficiently close to this singularity.

In the pinching of a Newtonian liquid free surface, the system spontaneously approaches a finite-time singularity, which offers a unique opportunity to observe the behavior of fluids with arbitrarily small length and time scales. This property and its universal character (insensitivity to both initial and boundary conditions) turn this problem into an ideal candidate to question our knowledge of fundamental aspects of fluid dynamics. Both theoretical \citep{PSS90,E93,P95,LS16,KWTB18} and experimental \citep{CMP09,VMHF14,CCTSHHLB15,PMHVV17} studies on the free surface pinch-off have traditionally considered the dependence of the free surface minimum radius, $R_{\textin{min}}$, with respect to the time to the pinching, $\tau$, as an indicator of the relevant forces arising next to the pinching spatiotemporal coordinate. For small viscous effects, the thinning of the liquid thread passes through an inertio-capillary regime characterized by the power law
\begin{equation}
\label{ebbb}
R_{\textin{min}}=A \left(\frac{\sigma}{\rho}\right)^{1/3} \tau^{2/3},
\end{equation}
where $\sigma$ and $\rho$ are the liquid surface tension and density, respectively \citep{KM83,E93}. The dimensionless prefactor $A$ can exhibit a complex, nonmonotonic behavior over many orders of magnitude in $\tau$. In fact, its asymptotic value $A\simeq 0.717$ is never reached because there are very long-lived transients, and then viscous effects take over \citep{DHHVRKEB18}.

The addition of surfactant confers a certain degree of complexity on Newtonian liquids, which may lead to unexpected behaviors during the pinch-off of their free surfaces. For instance, Marangoni stress can produce microthread cascades during the breakup of interfaces loaded with surfactants \citep{MB06}. It is still a subject of debate whether surfactants are convected away from the pinching region. In that case, the system would follow the self-similar dynamics of clean interfaces at times sufficiently close to the breakup \citep{TL02,CMP02,LSFB04,LFB06,LB07,CMP09,RABK09,CMCP11a,PMHVV17}. The persistence of a surfactant monolayer in the pinching of an interface potentially entails the appearance of several effects. The first and probably more obvious is the so-called solutocapillarity, i.e., the local reduction of the surface tension due to the presence of surface-active molecules \citep{RABK09,SPADBK12}. The other effect that has been accounted for is the Marangoni stress induced by the surface tension gradient due to uneven distribution of surfactant along the free surface \citep{SL90b,AB99,TL02,CMP02,MB06,HSYLBP08,JGS06,LFB06,DSXCS06,HM16b,KWTB18}. However, some other effects might be considered in the vicinity of the pinching region as well. Among them, the shear and dilatational surface viscosities have already been shown to affect considerably the breakup of pendant drops covered with insoluble (viscous) surfactants \citep{PMHVV17}.

SDS is one of the most commonly used surfactants in microfluidic experiments. The adsorption/desorption times of SDS are several orders of magnitude larger than the characteristic time of the breakup of free surfaces enclosing low-viscosity liquids. This allows one to regard SDS as an insoluble surfactant, which considerably simplifies the problem. Under the insolubility condition, bulk diffusion and adsorption/desorption processes can be ruled out. Due to its small molecular size, the SDS monolayer is assumed to exhibit a Newtonian behavior \citep{S60}. In addition, the sphere-to-rod transition of SDS micelles (and its associated viscoelastic behavior) does not take place unless some specific salt is added to the solution \citep{AKC03}. Therefore, viscoelastic effects are not expected to come up even for concentrations larger than the cmc.

Surface viscosities of small-size surfactant molecules, such as SDS, are believed not to affect the breakage of a pendant drop due to their small values. However, and as mentioned above, the surface-to-volume ratio diverges in the vicinity of the pinching region and, therefore, surface viscous effects can eventually dominate both inertia and viscous dissipation in the bulk of that region. In addition, the surface tension is bounded between the values corresponding to the clean free surface and the maximum packaging limit, while surface velocity can diverge at the pinch-off singularity. This suggests that surface viscous stresses (which are proportional to the surface velocity gradient) can become comparable with, or even greater than, Marangoni stress (which is proportional to surface tension gradient) in the pinching region for times sufficiently close to the breakup. One can hypothesize that surface viscous stresses can eventually have a measurable influence on the evolution of the free surface even for very low-viscosity surfactants. This work aims to test this hypothesis. The comparison between numerical simulations and experimental data will allow us to determine upper bounds for both the shear and dilatational viscosities of SDS.

\section{Theoretical model}
\label{sec2}

\begin{figure}[h]
\vcenteredhbox{\resizebox{0.15\textwidth}{!}{\includegraphics{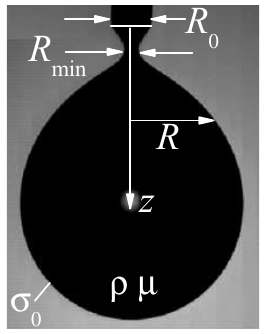}}}
\caption{Image of a pendant drop in the experiments right before its breakup.}
\label{sketch}
\end{figure}

Consider a liquid drop of density $\rho$ and viscosity $\mu$ hanging on a vertical capillary (needle) of radius $R_0$ due to the action of the (equilibrium) surface tension $\sigma_0$ (Fig.\ \ref{sketch}). In this section, all the variables are made dimensionless with the needle radius $R_0$, the inertio-capillary time $t_0=(\rho R_0^3/\sigma_0)^{1/2}$, the inertio-capillary velocity $v_0=R_0/t_0$, and the capillary pressure $\sigma_0/R_0$. The velocity ${\bf v}({\bf r},t)$ and reduced pressure $p({\bf r},t)$ fields are calculated from the continuity and Navier-Stokes equations
\begin{equation}
{\boldsymbol \nabla}\cdot {\bf v}=0,
\end{equation}
\begin{equation}
\frac{\partial {\bf v}}{\partial t}+{\bf v}\cdot {\boldsymbol \nabla}{\bf v}=-{\boldsymbol \nabla}p+{\boldsymbol \nabla}\cdot {\bf T},
\end{equation}
respectively, where ${\bf T}=\text{Oh}[{\boldsymbol \nabla}{\bf v}+({\boldsymbol \nabla}{\bf v})^T]$ is the viscous stress tensor, and $\text{Oh}=\mu(\rho\sigma_0 R_0)^{-1/2}$ is the volumetric Ohnesorge number. These equations are integrated over the liquid domain of (dimensionless) volume $V$ considering the non-slip boundary condition at the solid surface, the anchorage condition at the needle edge, and the kinematic compatibility condition at the free surface.

Neglecting the dynamic effects of the surrounding gas, the balance of normal stresses at the free surface yields \cite{LH98}
\begin{equation}
-p+B\, z+{\bf n}\cdot {\bf T}\cdot {\bf n}=[\hat{\sigma}+(\text{Oh}_2^S-\text{Oh}_1^S)\boldsymbol{\nabla}^S\cdot\mathbf{v}^S]\kappa+2\text{Oh}_1^S[\kappa_1(\boldsymbol{\nabla}^S\mathbf{v}^S)_{11}+\kappa_2(\boldsymbol{\nabla}^S\mathbf{v}^S)_{22}]
,\label{NormalStress1}
\end{equation}
where $B=\rho g R_0^2/\sigma_0$ is the Bond number, $g$ the gravitational acceleration, ${\bf n}$ the unit outward normal vector, $\widehat{\sigma}\equiv\sigma/\sigma_0$ is the ratio of the local value $\sigma$ of the surface tension to its equilibrium value $\sigma_0$, $\text{Oh}_{1,2}^S=\mu_{1,2}^S(\rho\sigma_0 R_0^3)^{-1/2}$ are the superficial Ohnesorge numbers defined in terms of the surface shear and dilatational viscosities $\mu_1^S$ and $\mu_2^S$, respectively, $\bnabla^ S$ the tangential intrinsic gradient along the free surface, $\mathbf{v}^S(z,t)$ the (two-dimensional) tangential velocity to the free surface, $\kappa=\kappa_1+\kappa_2$ (twice) the mean curvature of the free surface, $\kappa_1$ and $\kappa_2$ the curvatures along the meridians and parallels in the inward normal direction, respectively, and $(\boldsymbol{\nabla}^S\mathbf{v}^S)_{11}$ and $(\boldsymbol{\nabla}^S\mathbf{v}^S)_{22}$ the diagonal elements of $\boldsymbol{\nabla}^S\mathbf{v}^S$ along the meridians and the parallels, respectively.

In addition, the balance of tangential stresses leads to
\begin{equation}
{\bf t}\cdot {\bf T}\cdot {\bf n}={\bf t}\cdot \btau^S,
\end{equation}
where ${\bf t}$ is the unit vector tangential to the free surface meridians, and
\begin{eqnarray}
\btau^S=\bnabla^S\widehat{\sigma}+\bnabla^S\cdot\{ \text{Oh}_1^S[\bnabla^S\mathbf{v}^S+(\bnabla^S\mathbf{v}^S)^\top]\}+\bnabla^S[(\text{Oh}_2^S-\text{Oh}_1^S)\bnabla^S\cdot\mathbf{v}^S],
\label{stress}
\end{eqnarray}
is the surface stress tensor.

The surface viscosities are expected to depend on the surfactant surface concentration. For the sake of simplicity, we assume the linear relationships $\mu_{1,2}^S=\mu_{1,2}^{S*}\widehat{\Gamma}/\widehat{\Gamma}_{\textin{cmc}}$, where $\mu_{1,2}^{S*}$ are the surfactant viscosities at the cmc. In addition, $\widehat{\Gamma}\equiv \Gamma/\Gamma_0$ and $\widehat{\Gamma}_{\textin{cmc}}\equiv\Gamma_{\textin{cmc}}/\Gamma_0$, where $\Gamma$ and $\Gamma_{\textin{cmc}}$ are the surfactant surface concentration and its value at the cmc, respectively, both in terms of the equilibrium value $\Gamma_0$. Therefore,
\begin{equation}
\text{Oh}_{1,2}^S=\text{Oh}_{1,2}^{S*} \frac{\widehat{\Gamma}}{\widehat{\Gamma}_{\textin{cmc}}},
\end{equation}
where $\text{Oh}_{1,2}^{S*}=\mu^{S*}_{1,2}(\rho\sigma_0 R_0^3)^{-1/2}$ are the superficial Ohnesorge numbers at the cmc.

To calculate the surfactant surface concentration, we take into account that the droplet breakup time is much smaller than the characteristic adsorption-desorption times (see Sec.\ \ref{sec3}), and, therefore, surfactant solubility can be neglected over the breakup process. In this case, one must consider the equation governing the surfactant transport on the free surface:
\begin{equation}
\label{conser}
\frac{\partial \widehat{\Gamma}}{\partial t}+{\bnabla}^S\cdot (\widehat{\Gamma}{\bf v})=\frac{1}{\text{Pe}^S}\, {\bnabla}^{S2}\widehat{\Gamma},
\end{equation}
where Pe$^S$=$R_0^2/(t_0 {\cal D}^S)$ and ${\cal D}^S$ are the surface Peclet number and diffusion coefficient, respectively. The equation of state $\widehat{\sigma}(\widehat{\Gamma})$ is obtained from experimental data as explained below.

The above theoretical model is numerically solved by mapping the time-dependent liquid region onto a fixed numerical domain through a coordinate transformation. The hydrodynamic equations are spatially discretized with the Chebyshev spectral collocation technique, and an implicit time advancement is performed using second-order backward finite differences \cite{HM16a}. To deal with the free surface overturning taking place right before the droplet breakup, a quasi-elliptic transformation \citep{DT03} was applied to generate the mesh. To trigger the pendant drop breakup process, a very small force was applied to a stable shape with a volume just below the critical one. This perturbation was expected to affect neither the pendant drop dynamics close to the free-surface pinch-off nor the formation of the satellite droplet. The time-dependent mapping of the physical domain does not allow the algorithm to surpass the free surface pinch-off, and therefore the evolution of the satellite droplet cannot be analyzed.

\section{Experimental method}
\label{sec3}

\begin{figure}[h]
\begin{tabular}{lr}
\vcenteredhbox{\resizebox{0.45\textwidth}{!}{\includegraphics{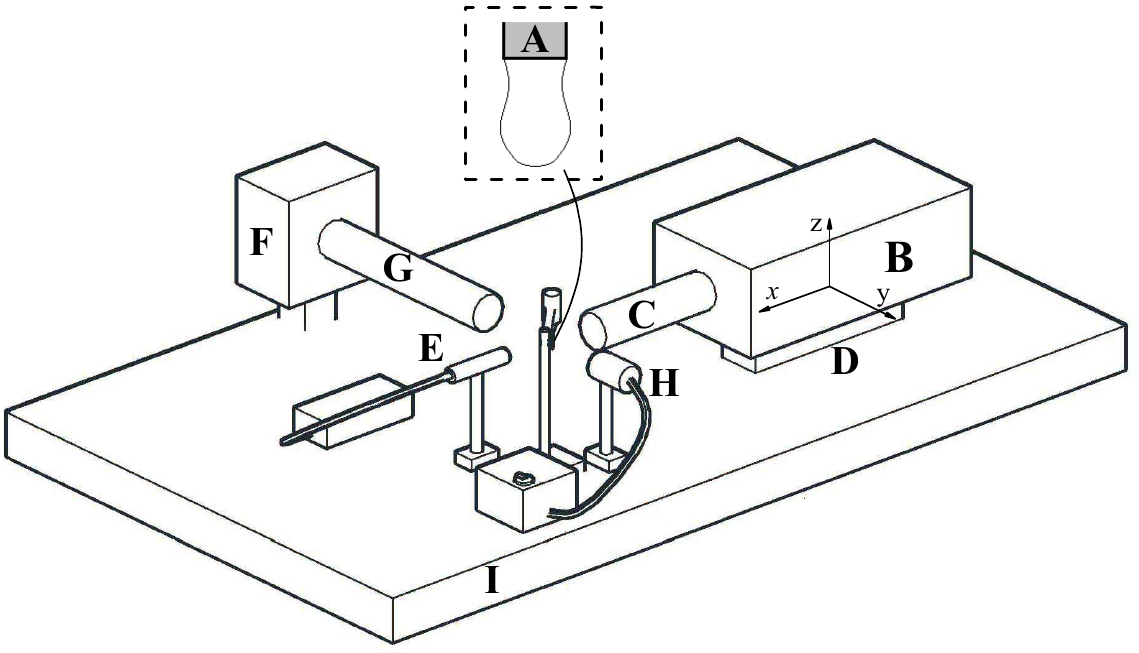}}}&
\vcenteredhbox{\resizebox{0.3\textwidth}{!}{\includegraphics{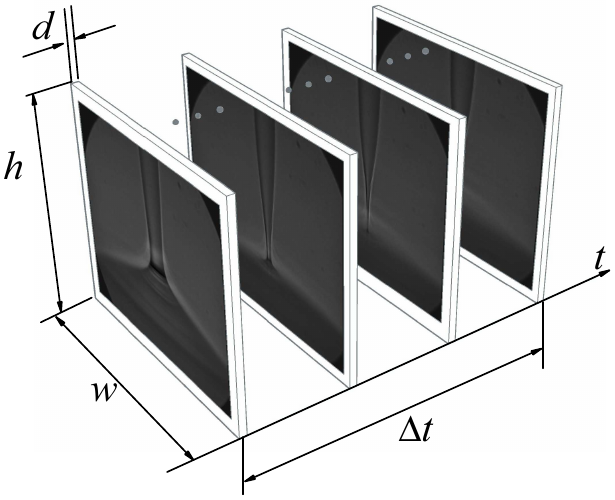}}}
\end{tabular}
\caption{(Left) Experimental setup: feeding capillary (A), ultra-high speed video camera (B), optical lenses (C), triaxial translation stage (D), laser (E), optical trigger (F), optical lenses (G), white backlight (H), and anti-vibration isolation system (I). (Right) Spatio-temporal hypervolume analyzed in the experiment: image width $w=94$ $\mu$m, height $h=78$ $\mu$m, depth of field $d=0.48$ $\mu$m and time $\Delta t=36$ $\mu$s
elapsed during the experiment.}
\label{setup}
\end{figure}

In the experimental setup (Fig.\ \ref{setup}), a cylindrical feeding capillary (A) $R_0=115$ $\mu$m in outer radius was placed vertically. To analyze the role of the capillary size, we also conducted experiments with $R_0=205$ $\mu$m. A pendant droplet was formed by injecting the liquid at a constant flow rate with a syringe pump (Harvard Apparatus PHD 4400) connected to a stepping motor. We used a high-precision orientation system and a translation stage to ensure the correct position and alignment of the feeding capillary. Digital images of the drop were taken using an ultra-high-speed video camera ({\sc Kirana}-5M) (B) equipped with optical lenses (an Optem HR 50X magnification zoom-objective and a NAVITAR 12X set of lenses) (C). As explained below, the images were acquired either at $5\times 10^6$ fps with a magnification 101.7 nm/pixel or at $5\times 10^5$ fps with a magnification 156 nm/pixel. The camera could be displaced both horizontally and vertically using a triaxial translation stage (D) with one of its horizontal axes (axis $x$) motorized (THORLABS Z825B) and controlled by the computer, which allowed as to set the droplet-to-camera distance with an error smaller than 29 nm. The camera was illuminated with a laser (SI-LUX 640, {\sc Specialised Imaging}) (E) synchronized with the camera, which reduced the effective exposure time down to 100 ns. The camera was triggered by an optical trigger (SI-OT3, {\sc Specialised Imaging}) (F) equipped with optical lenses (G) and illuminated with cold white backlight (H). All these elements were mounted on an optical table with a pneumatic anti-vibration isolation system (I) to damp the vibrations coming from the building.

In the experiment, a pendant droplet hanging on the feeding capillary was inflated by injecting the liquid at 1 ml/h. The triple contact lines anchored to the outer edge of the capillary. The drop reached its maximum volume stability limit after around 20 s. We analyzed images of the quasi-static process with the Theoretical Image Fitting Analysis (TIFA) \citep{CBMN04} method to verify that the surface tension right before the droplet breakup was the same (within the experimental uncertainty) as that measured at equilibrium. In this way, one can ensure that the surfactant surface concentration corresponded to the prescribed volumetric concentration at equilibrium. This conclusion can be anticipated from the fact that the characteristic surfactant adsorption process is much smaller than the droplet inflation time.

When the maximum volume stability limit was reached, the droplet broke up spontaneously. We recorded 180 images at $5\times 10^6$ fps of the final stage of the breakup process within a spatial window $94\times 78$ $\mu$m. This experiment was repeated several times to assess the degree of reproducibility of the experimental results. The flow rate at which the pendant droplet is inflated was reduced down to 0.1 ml/h to verify that this parameter did not affect the final stage of the breakup process. Besides, 180 images of a spatial window $144\times 120$ $\mu$m were taken at $5\times 10^5$ fps to describe the process on a larger scale.

\begin{figure}[h]
\begin{tabular}{lr}
\vcenteredhbox{\resizebox{0.3\textwidth}{!}{\includegraphics{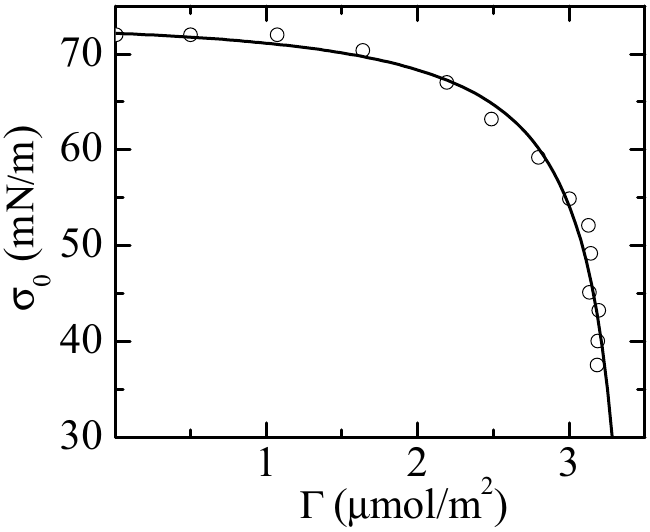}}}
\end{tabular}
\caption{Experimental values of the (equilibrium) surface tension $\sigma_0$ versus the surface surfactant concentration $\Gamma$ for SDS in DIW (symbols) \citep{TMS70}. The line corresponds to the fit (\ref{fit}) to those values.}
\label{tas}
\end{figure}

We selected SDS in deionized water (DIW) because it is a solution widely used in experiments and very well characterized. The dependence of the (equilibrium) surface tension with respect to the surface surfactant concentration $\Gamma$ has been determined from direct measurements (Fig.\ \ref{tas}) \citep{TMS70}. We use the fit
\begin{equation}
\label{fit}
\sigma_0=10^3\frac{-17.94\,\Gamma+60.76}{\Gamma^2-240.9\,\Gamma+841.8}
\end{equation}
to that experimental data in our simulations. In this equation, $\sigma_0$ and $\Gamma$ are measured in mN/m and $\mu$mol/m$^2$, respectively. It should be noted that there is no theoretical justification for the above equation of state. It simply represents an accurate approximation for the numerical simulations. Other equations may be equally valid for our purposes.

Table \ref{tab2} shows some physical properties of SDS in DIW. The shear $\mu_1^{S*}$ and dilatational $\mu_2^{S*}$ surface viscosities of aqueous solutions of SDS at the cmc have been widely measured with different methods over the last decades. \citet{ZNMLDMTS14} reported the surface shear viscosity to be below $10^{-8}$ Pa\,s\,m (the sensitivity limit of their technique). Other authors have measured values up to five orders of magnitude higher than that upper bound \citep{MK12,LD03}.

\begin{table*}
\begin{tabular}{|c|c|c|c|c|c|c|c|}
\hline
$\mu_1^{S*}$ (Pa\,s\,m) \citep{ZNMLDMTS14} & $\mu_2^{S*}$ (Pa\,s\,m) \citep{MK12} & ${\cal D}^S$ (m$^2$/s) \citep{MK12} & $t_a$ (ms) \citep{RABK09} & $t_d$ (ms) \citep{MK12} & $\Gamma_{\textin{cmc}}$ ($\mu$mol m$^{-2}$ & $N_{\textin{agg}}$ \citep{CDA08} & $R_{\textin{mic}}$ (nm) \citep{CDA08}\\
\hline
$<10^{-8}$ & $10^{-7}$--$10^{-9}$ & $8\times 10^{-10}$ & 100 & 169.5 & 3.19 & 61 & 1.72\\
\hline
\end{tabular}
\caption{Physical properties of SDS in DIW: superficial viscosities $\mu_{1,2}^{S*}$, , surfactant surface diffusivity ${\cal D}^S$, adsorption $t_a$ and desorption $t_d$ time, aggregation number $N_{\textin{agg}}$, and micelle radius $R_{\textin{mic}}$.}
\label{tab2}
\end{table*}

Table \ref{tab3} shows the values of the superficial Ohnesorge numbers, Boussinesq numbers $\text{Bq}_{1,2}=\mu_{1,2}^S/(\mu \ell_c)$, and surface Peclet number. The superficial Ohnesorge numbers are much smaller than the volumetric one, $\text{Oh}\simeq 0.02$, which indicates that the superficial viscosities play no significant role on a scale given by the feeding capillary radius $R_0$. The Boussinesq numbers are defined in terms of the characteristic length $\ell_c\equiv 1$ $\mu$m of the pinching region (see Sec.\ \ref{sec4}). Due to the smallness of this length, superficial viscous stresses may become comparable with the bulk ones, and, therefore, may produce a measurable effect on that scale. The value of the Peclet number indicates that surfactant surface diffusion is negligible at the beginning of the droplet breakup. The Peclet number defined in terms of $\ell_c$ and the corresponding capillary time $(\rho\ell_c^3/\sigma_0)^{1/2}$ takes values of the order of $10^3-10^4$. Therefore, one can expect surface diffusion to play a secondary role on that scale too.

\begin{table*}
\begin{tabular}{|c|c|c|c|c|}
\hline
Oh$_1^S$ & Oh$_2^S$ & $\text{Bq}_1$ & $\text{Bq}_2$ & $\text{Pe}^S$\\
\hline
$<9.35\times 10^{-4}$ & $9.35\times 10^{-3}$--$9.35\times 10^{-5}$ & $<1.41$ & 14.1--0.14 & $7.73\times 10^{4}$\\
\hline
\end{tabular}
\caption{Dimensionless numbers calculated from the physical properties of SDS in DIW (Table \ref{tab2}): interfacial Ohnesorge numbers Oh$_{1,2}^S$, Boussinesq numbers $\text{Bq}_{1,2}$, and surface Peclet number $\text{Pe}^S$.}
\label{tab3}
\end{table*}

\section{Results}
\label{sec4}

\begin{figure}[h]
\vcenteredhbox{\resizebox{0.45\textwidth}{!}{\includegraphics{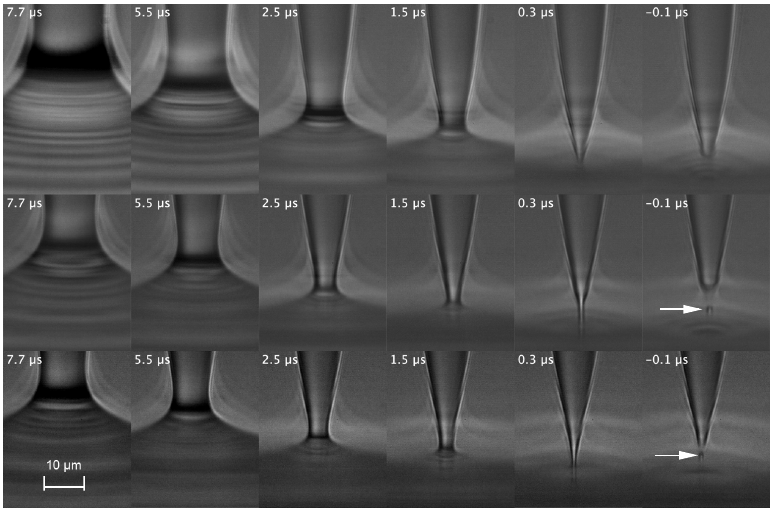}}}
\caption{(From top to bottom) Pinch-off of a drop of DIW, DIW+SDS 0.8cmc, and DIW+SDS 2cmc. The labels indicate the time to the pinching with an error of $\pm$100 ns. The arrows point to the subsatellite droplets.}
\label{imagesW}
\end{figure}

Figure \ref{imagesW} shows images of the pinch-off of a drop of DIW, DIW+SDS 0.8cmc, and DIW+SDS 2cmc. A microthread forms next to the pinching point when the surfactant is added. The breakup of that microthread produces a tiny subsatellite droplet 1-2 $\mu$m in diameter. This droplet is significantly smaller than that observed in previous experiments with 5-cSt silicone oil in the absence of surfactant, which seems to confirm that the silicone oil subsatellite droplet was formed by viscoelastic effects \citep{RPVHM19}.

\begin{figure}[h]
\vcenteredhbox{\resizebox{0.3\textwidth}{!}{\includegraphics{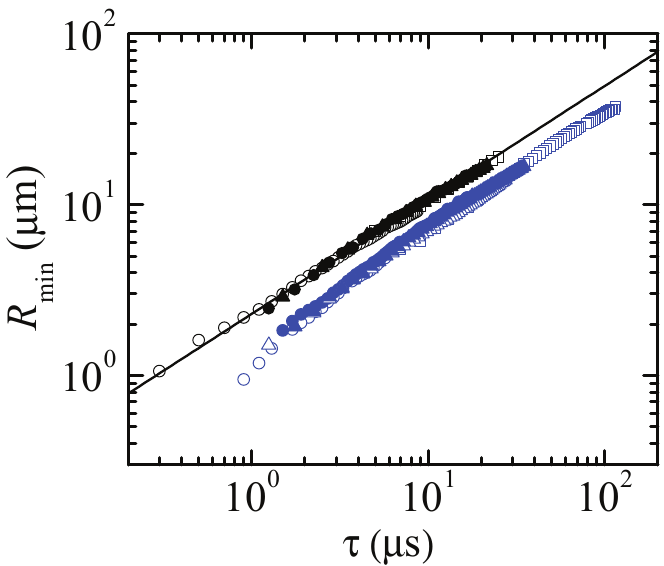}}}
\caption{$R_{\textin{min}}(\tau)$ for the breakup of a pendant drop of DIW and DIW+SDS 0.8cmc. The black and blue symbols are the experimental data for DIW and DIW+SDS 0.8cmc, respectively. The different symbols correspond to experiments visualized with different magnifications and recording speeds. The open and solid symbols correspond to experiments conducted with a cylindrical feeding capillary $R_0=115$ and $205$ $\mu$m in radius, respectively. The solid line is the power law (\ref{ebbb}) with $A\simeq 0.55$.}
\label{universal}
\end{figure}

Figure \ref{universal} shows the free surface minimum radius, $R_{\textin{min}}$, as a function of the time to the pinching, $\tau$, for experiments conducted with two feeding capillary radii. The agreement among the results obtained for the same liquid shows both the high reproducibility of the experiments and the universal character (independency from $R_0$) of $R_{\textin{min}}(\tau)$ for the analyzed time interval. In fact, the differences between the results obtained with $R_0=115$ and $205$ $\mu$m are smaller than the effect attributed to the surface viscosities, as will be described below. The results for DIW follow the scaling law (\ref{ebbb}) with $A\simeq 0.55$.

\begin{figure}[h]
\vcenteredhbox{\resizebox{0.3\textwidth}{!}{\includegraphics{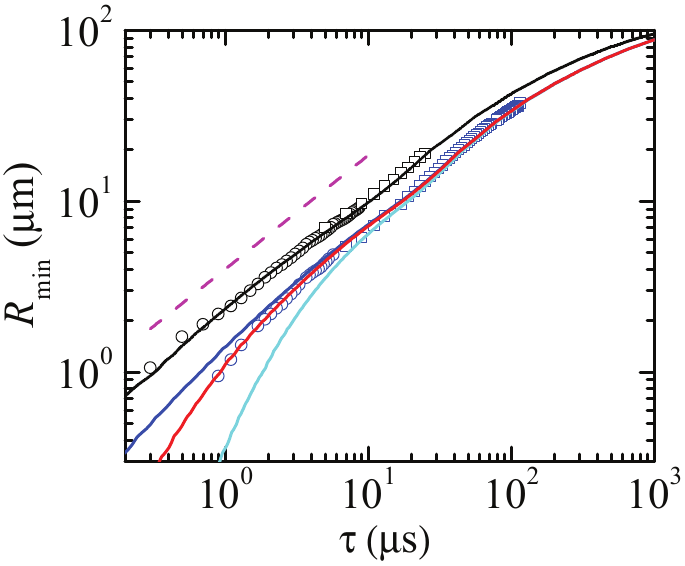}}}\vcenteredhbox{\resizebox{0.3\textwidth}{!}{\includegraphics{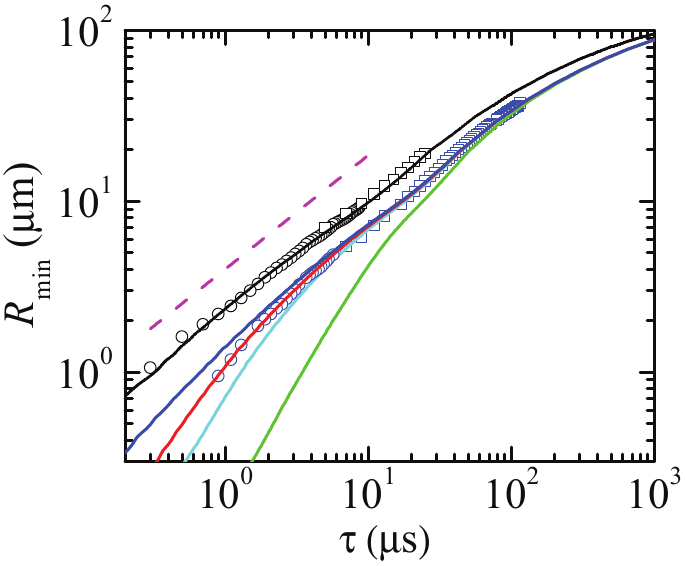}}}
\caption{$R_{\textin{min}}(\tau)$ for the breakup of a pendant drop of DIW and DIW+SDS 0.8cmc. The black and blue symbols are the experimental data for DIW and DIW+SDS 0.8cmc, respectively. The different symbols correspond to experiments visualized with different magnifications. The black solid line and magenta dashed line correspond to the simulation and the power law $R_{\textin{min}}(\tau)\sim \tau^{2/3}$ for DIW, respectively. (Left) The colored solid lines correspond to simulations of DIW+SDS 0.8cmc for $\mu_2^{S*}=0$ and $\mu_1^{S*}=0$ (blue), $5 \times 10^{-10}$ (red), and $3.5 \times 10^{-9}$ Pa\,s\,m (cyan). (Right) The colored solid lines correspond to simulations of DIW+SDS 0.8cmc for $\mu_1^{S*}=0$ and $\mu_2^{S*}=0$ (blue), $3.5 \times 10^{-9}$ (red), $10^{-8}$ (cyan), and $10^{-7}$ Pa\,s\,m (green).}
\label{W08}
\end{figure}

\begin{figure}[h]
\vcenteredhbox{\resizebox{0.3\textwidth}{!}{\includegraphics{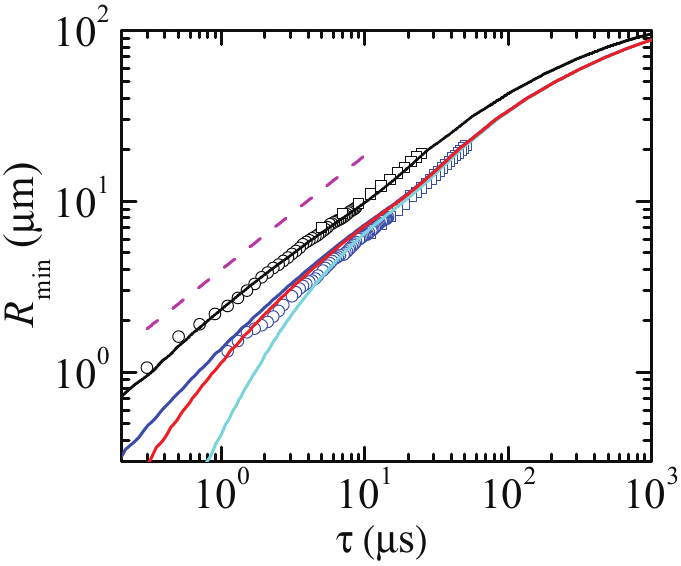}}}\vcenteredhbox{\resizebox{0.3\textwidth}{!}{\includegraphics{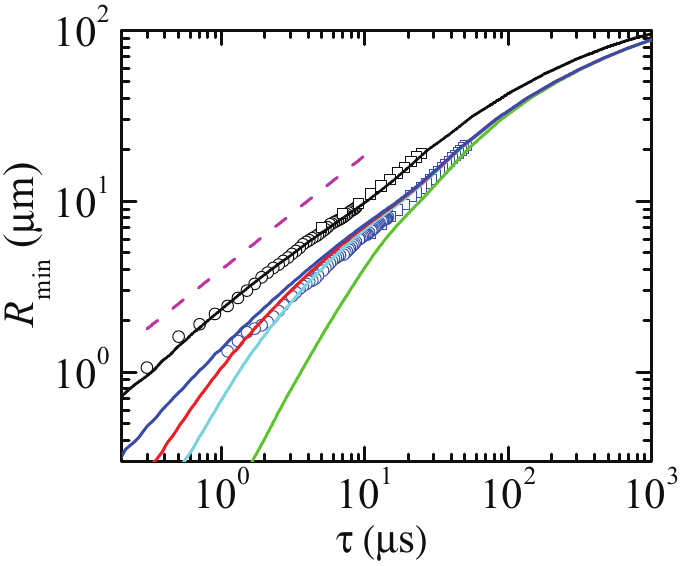}}}
\caption{$R_{\textin{min}}(\tau)$ for the breakup of a pendant drop of DIW and DIW+SDS 2cmc. The black and blue symbols are the experimental data for DIW and DIW+SDS 2cmc, respectively. The different symbols correspond to experiments visualized with different magnifications. The black solid line and magenta dashed line correspond to the simulation and the power law $R_{\textin{min}}(\tau)\sim \tau^{2/3}$ for DIW, respectively. (Left) The colored solid lines correspond to simulations of DIW+SDS 2cmc for $\mu_2^{S*}=0$ and $\mu_1^{S*}=0$ (blue), $5 \times 10^{-10}$ (red), and $3.5 \times 10^{-9}$ Pa\,s\,m (cyan). (Right) The colored solid lines correspond to simulations of DIW+SDS 2cmc for $\mu_1^{S*}=0$ and $\mu_2^{S*}=0$ (blue), $3.5 \times 10^{-9}$ (red), $10^{-8}$ (cyan), and $10^{-7}$ Pa\,s\,m (green).}
\label{W2}
\end{figure}

As can be seen in Figs.\ \ref{W08} and \ref{W2}, there is a remarkable agreement between the experiments and numerical simulations for the pure DIW case for times to the pinching as small as $\sim 300$ ns, which constitutes a stringent validation of both experiments and simulations. When SDS is dissolved in water, it creates a monolayer which substantially alters the pinch-off dynamics. The function $R_{\textin{min}}(\tau)$ takes smaller values than in the pure DIW case over the entire process due to the reduction of the surface tension. More interestingly, if only solutocapillarity and Marangoni convection are considered in the numerical simulations (blue solid lines), there is a measurable deviation with respect to the experimental results for $R_{\textin{min}}(\tau)\lesssim 5$ $\mu$m. Specifically, the free surface in the experiment evolves towards its pinching slower than in the numerical simulation. We added surface viscous stresses to the simulation to reproduce the entire range of experimental data. To this end, we set to zero one of the surface viscosities and modulated the other. In this way, one can establish upper bounds of both the shear $\mu_1^{S*}$ and extensional $\mu_2^{S*}$ viscosity.

The experimental results can be reproduced for $\mu_1^{S*}=5 \times 10^{-10}$ Pa\,s\,m and $\mu_2^{S*}=0$ (see Figs.\ \ref{W08}-left and \ref{W2}-left). This upper bound is consistent with the results obtained by \citet{ZNMLDMTS14}, who concluded that the surface shear viscosity of SDS in DIW must take values below $10^{-8}$ Pa\,s\,m (the sensitivity limit of their technique). The experimental results can also be reproduced for $\mu_1^{S*}=0$ and $\mu_2^{S*}=3.5 \times 10^{-9}$ Pa\,s\,m (Figs.\ \ref{W08}-right and \ref{W2}-right). There are significant deviations when other values of $\mu_2^{S*}$ found in the literature are considered \citep{MK12}. The optimum value of the shear viscosity is one order of magnitude smaller than that of the dilatational viscosity, which suggests that shear viscous stresses have a greater effect on the pinching than dilatational ones for the same value of the corresponding surface viscosities. In fact, when the surface shear viscosity takes the value of the dilatational viscosity ($\mu_1^{S*}=3.5 \times 10^{-9}$ Pa\,s\,m, $\mu_2^{S*}=0$) the numerical curve (cyan solid line in Figs.\ \ref{W08}-left and \ref{W2}-left) significantly deviates from the experimental one. The relative importance of the shear and dilatational viscosities can be explained in terms of the equivalence between the corresponding terms in the 1D approximation discussed in Sec.\ \ref{sec2}. The agreement achieved for DIW+SDS 2cmc is slightly worse than that obtained for DIW+SDS 0.8cmc probably because the experimental surface tension values are less accurate for concentrations larger than the cmc (see Fig.\ \ref{tas}).

\begin{figure}[h]
\vcenteredhbox{\resizebox{0.35\textwidth}{!}{\includegraphics{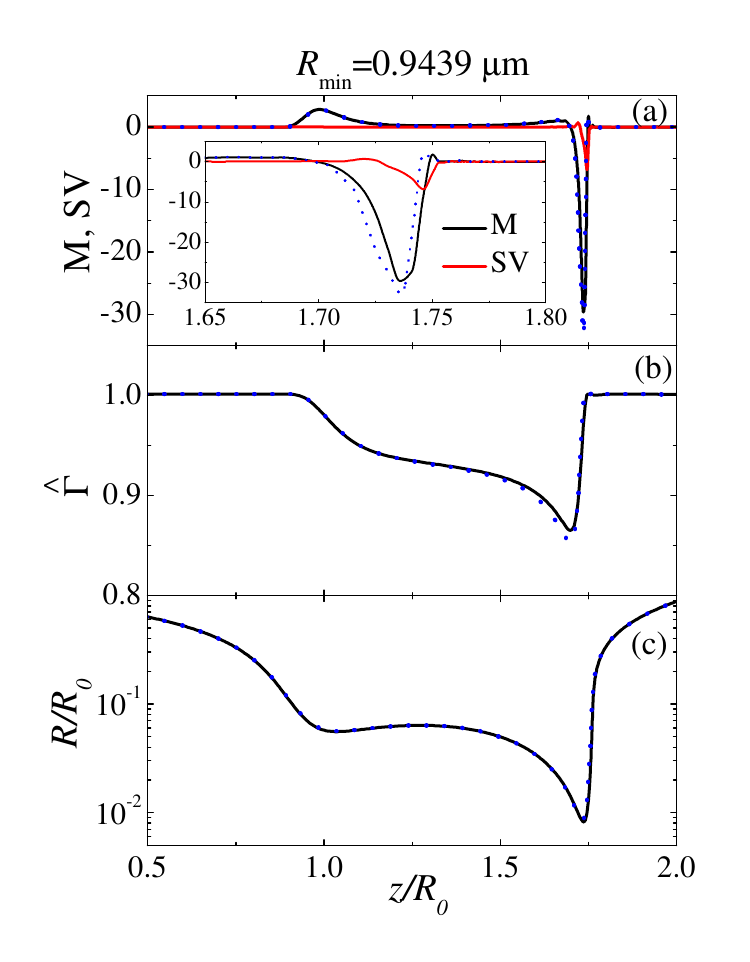}}}\vcenteredhbox{\resizebox{0.35\textwidth}{!}{\includegraphics{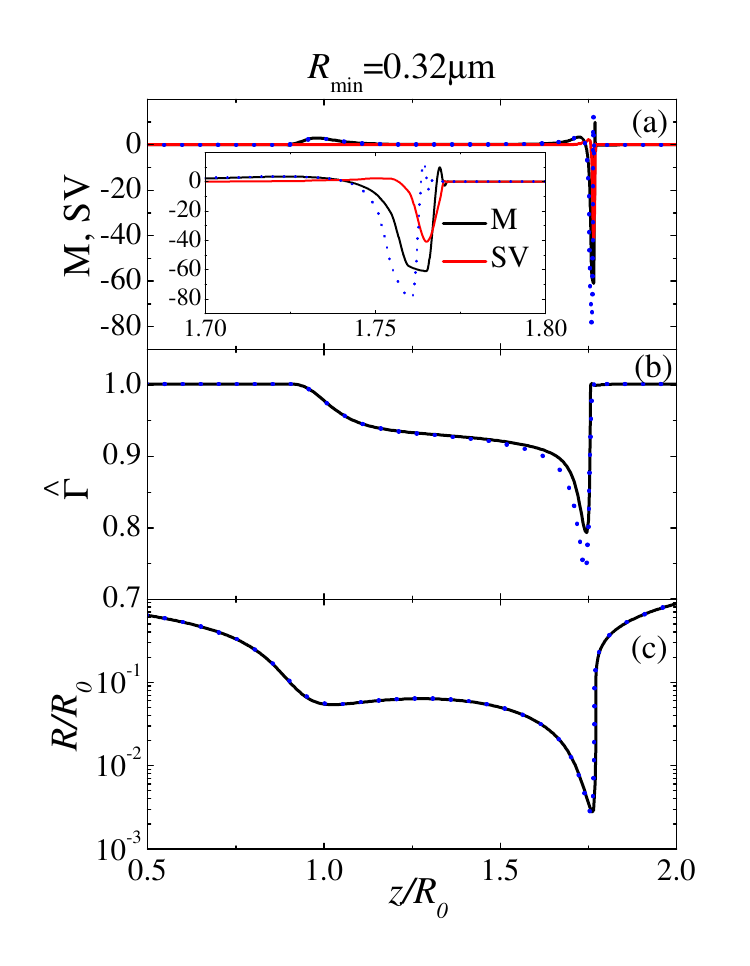}}}
\caption{Axial distribution of the Marangoni stress (M) and tangential shear viscous stress (SV) (a), surfactant surface concentration (b,) and free surface radius (c) for DIW+SDS 0.8cmc. The solid lines are the results for $\{\mu_1^{S*}=5 \times 10^{-10}$ Pa\,s\,m, $\mu_2^{S*}=0\}$, while the dotted lines correspond to $\mu_1^{S*}=\mu_2^{S*}=0$. The dotted lines show the results for $\mu_1^{S*}=\mu_2^{S*}=0$ (in the left-hand graphs, $R_{\textin{min}}=0.9836$ $\mu$m for $\mu_1^{S*}=\mu_2^{S*}=0$).}
\label{distributionshear}
\end{figure}

\begin{figure}[h]
\vcenteredhbox{\resizebox{0.35\textwidth}{!}{\includegraphics{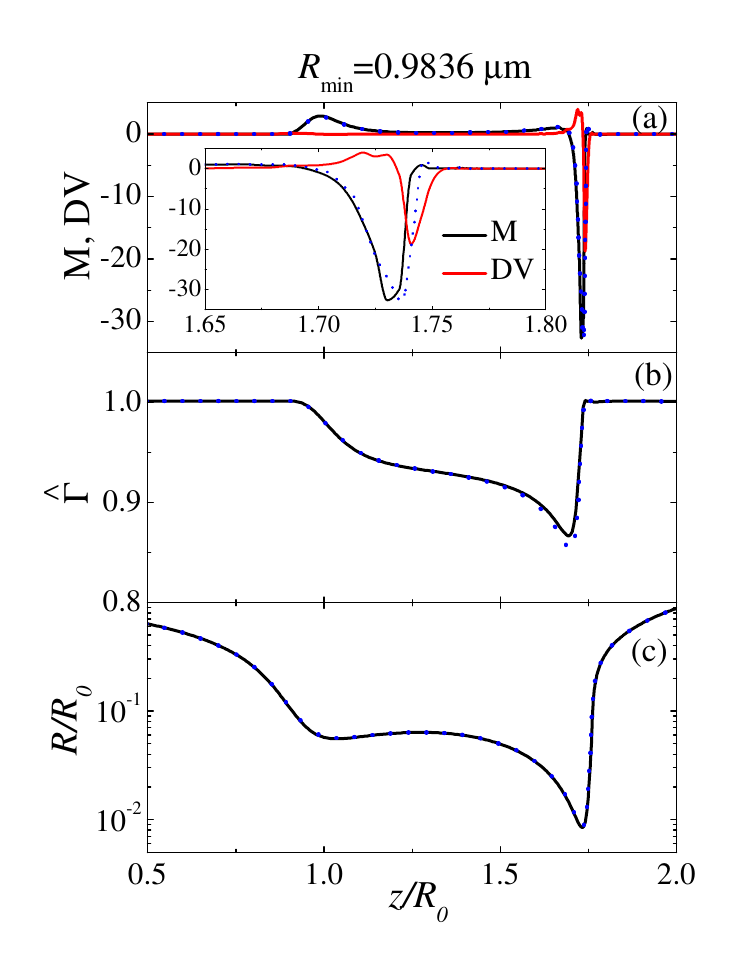}}}\vcenteredhbox{\resizebox{0.35\textwidth}{!}{\includegraphics{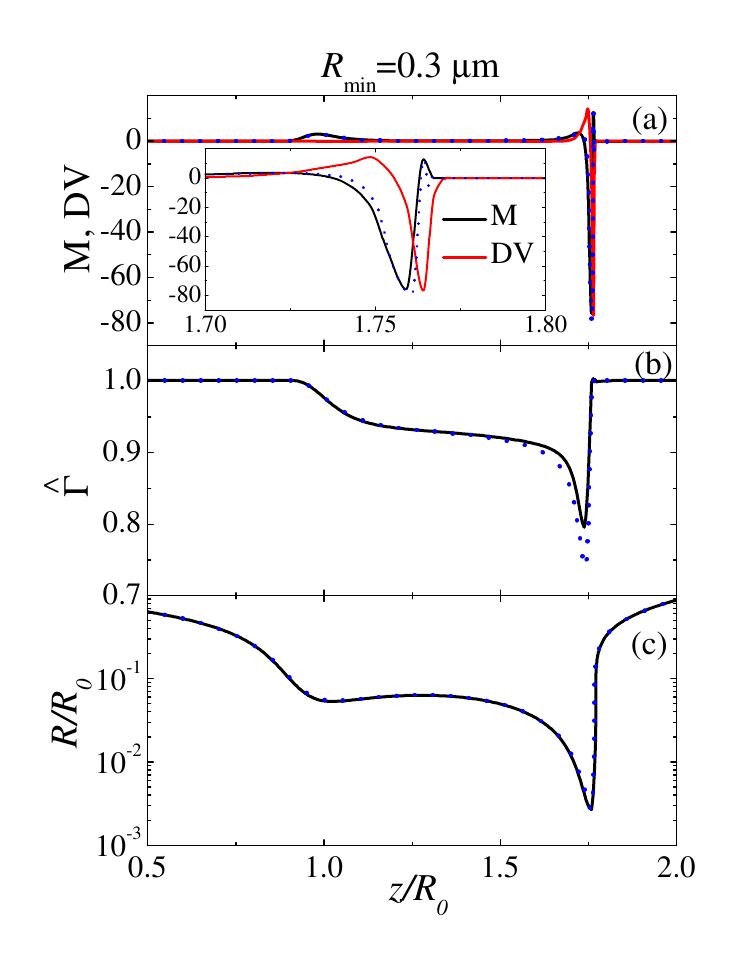}}}
\caption{Axial distribution of the Marangoni stress (M) and tangential dilatational viscous stress (DV) (a), surfactant surface concentration (b,) and free surface radius (c) for DIW+SDS 0.8cmc. The solid lines are the results for $\{\mu_1^{S*}=0$, $\mu_2^{S*}=3.5\times 10^{-9}$ Pa\,s\,m$\}$, while the dotted lines correspond to $\mu_1^{S*}=\mu_2^{S*}=0$. The dotted line show the results for $\mu_1^{S*}=\mu_2^{S*}=0$ (in the right-hand graphs, $R_{\textin{min}}=0.32$ $\mu$m for $\mu_1^{S*}=\mu_2^{S*}=0$).}
\label{distributiondilatational}
\end{figure}

Equation (\ref{stress}) shows the competition between the Marangoni stress, $\text{M}\equiv {\bf t}\cdot\bnabla^S\widehat{\sigma}$, and the tangential projection of the surface viscous stress,
\begin{eqnarray}
\text{SV}\equiv {\bf t}\cdot\left[\bnabla^S\cdot\{ \text{Oh}_1^S[\bnabla^S\mathbf{v}^S+(\bnabla^S\mathbf{v}^S)^\top]\}-\bnabla^S(\text{Oh}_1^S\bnabla^S\cdot\mathbf{v}^S)\right]\quad \text{and} \quad \text{DV}\equiv  {\bf t}\cdot\left[\bnabla^S(\text{Oh}_2^S\bnabla^S\cdot\mathbf{v}^S)\right],
\label{svis}
\end{eqnarray}
where SV and DV are the (dimensionless) contributions associated with the shear and dilatational surface viscosities, respectively. Figures \ref{distributionshear} and \ref{distributiondilatational} show the axial distribution of the tangential stresses, surfactant surface concentration, and free surface radius at a given instant of the droplet evolution. In Fig.\ \ref{distributionshear}, we compare the solution for $\mu_1^{S*}=\mu_2^{S*}=0$ with that for $\mu_2^{S*}=0$ and the optimum value of the shear surface viscosity determined from Fig.\ \ref{W08}-left, $\mu_1^{S*}=5 \times 10^{-10}$ Pa\,s\,m. The same comparison is presented in Fig.\ \ref{distributiondilatational} but for $\mu_1^{S*}=0$ and the optimum value of the dilatational surface viscosity determined from Fig.\ \ref{W08}-right, $\mu_2^{S*}=3.5\times 10^{-9}$ Pa\,s\,m. The instants were selected so that $R_{\textin{min}}$ took approximately the same value in the simulations with and without surface viscosities.

Consider the solution for $\{\mu_1^{S*}=5 \times 10^{-10}$ Pa\,s\,m, $\mu_2^{S*}=0\}$ (Fig.\ \ref{distributionshear}). For $R_{\textin{min}}=0.9439$ $\mu$m, the shear viscous stress is much smaller than the Marangoni stress over the entire free surface. As the minimum radius decreases, the relative importance of the shear viscosity increases. In fact, the maximum value of the shear viscous stress becomes comparable to that of the Marangoni stress for $R_{\textin{min}}=0.32$ $\mu$m. Small differences in the surfactant distribution arise for $R_{\textin{min}}\lesssim 0.32$ $\mu$m. The presence of shear viscosity slightly reduces the magnitude of the Marangoni stress. 

\begin{figure}[h]
\vcenteredhbox{\resizebox{0.35\textwidth}{!}{\includegraphics{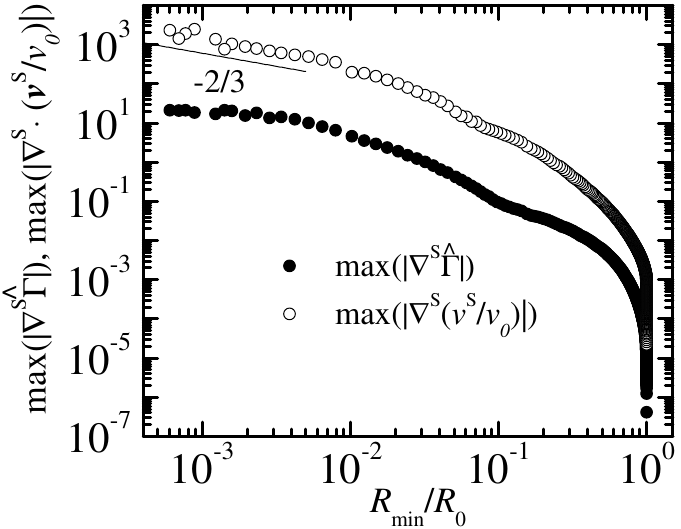}}}
\caption{Maximum values of the surfactant gradient, max($\boldsymbol{\nabla}^S\widehat{\Gamma}$) (solid symbols), and the surface velocity gradient, max($\boldsymbol{\nabla}^S\cdot{\bf v}^S$) (open symbols), for $\{\mu_1^{S*}=0$, $\mu_2^{S*}=3.5\times 10^{-9}$ Pa\,s\,m$\}$.}
\label{maximum}
\end{figure}

As mentioned in the Introduction, there is still a certain controversy about whether surfactants are convected away from the pinching region \citep{TL02,CMP02,LSFB04,LFB06,LB07,CMP09,RABK09,CMCP11a,PMHVV17}. Our results show that, when Marangoni and surface viscosity stresses are taken into account, the surfactant is not swept away from the thread neck in the time interval analyzed ($\widehat{\Gamma}\gtrsim 0.8$ in this region). These stresses operate in a different way but collaborate to keep the surfactant in the vicinity of the pinching point. Marangoni stress tries to restore the initial uniform surfactant concentration, while surface viscosity opposes to the variation of the surface velocity, and, therefore, to the extensional flow responsible for the surfactant depletion that would occur in the absence of Marangoni and viscous stresses. While the gradient of surfactant concentration remains bounded in the pinching region, the gradient of surface velocity continues to increase there (Fig.\ \ref{maximum}). This may explain why surface viscous stresses grow faster than Marangoni stress over the time interval analyzed. 

Interestingly, the free surface shape for $\mu_1^{S*}=\mu_2^{S*}=0$ is practically the same as that with the adjusted value of $\mu_1^{S*}$. This indicates that surface viscosity simply delays the time evolution of that shape. In fact, the values of the minimum radius obtained with and without surface viscosity significantly differ from each other when they are calculated at the same time to the pinching. For instance, $R_{\textin{min}}=0.32$ and 0.58 $\mu$m at $\tau\simeq 0.36$ $\mu$s for $\{\mu_1^{S*}=5 \times 10^{-10}$ Pa\,s\,m, $\mu_2^{S*}=0\}$ and $\mu_1^{S*}=\mu_2^{S*}=0$, respectively. However, the free surface shapes are practically the same if they are compared when the same value $R_{\textin{min}}=0.32$ $\mu$m of the minimum radius is reached. We can conclude that the surface viscosities of the SDS monolayer hardly alter the satellite droplet diameter and the amount of surfactant trapped in it. In this sense, solutocapillarity and Marangoni convection are the major factors associated with the surfactant \citep{KWTB18}.

\begin{figure}[h]
\vcenteredhbox{\resizebox{0.36\textwidth}{!}{\includegraphics{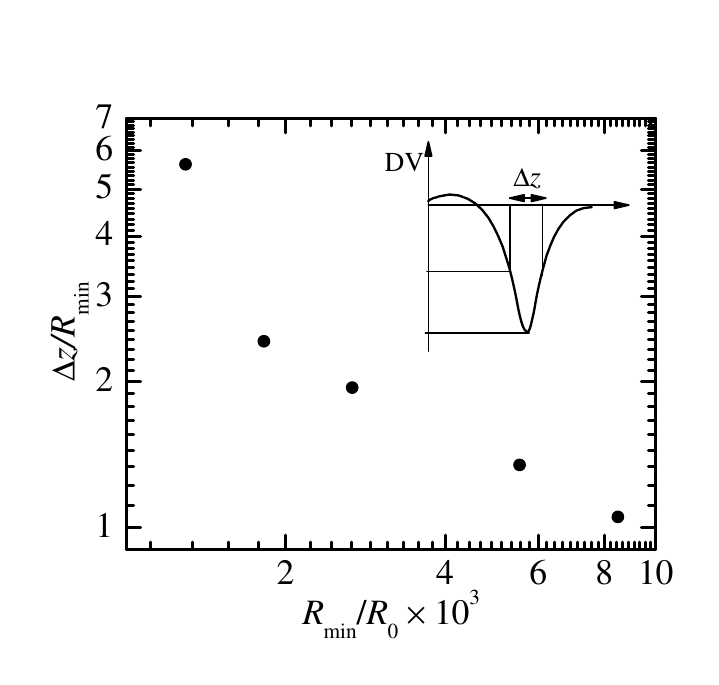}}}
\caption{Full width at half maximum, $\Delta z$, of the dilatational viscous stress as a function of the minimum radius $R_{\textin{min}}$ for DIW+SDS 0.8cmc with $\{\mu_1^{S*}=0$, $\mu_2^{S*}=3.5\times 10^{-9}$ Pa\,s\,m$\}$.}
\label{dz}
\end{figure}

Similar conclusions can be drawn from the numerical simulation conducted for $\{\mu_1^{S*}=0$, $\mu_2^{S*}=3.5\times 10^{-9}$ Pa\,s\,m$\}$ (Fig.\ \ref{distributiondilatational}). In this case, the dilatational viscous stress exhibits a noticeable maximum near the free surface neck. The full width at half maximum, $\Delta z$, measured in terms of the minimum radius, $R_{\textin{min}}$, sharply increases as the droplet approaches its breakup (Fig.\ \ref{dz}), which indicates that the importance of the dilatational viscous stress increases with time.

\begin{figure}[h]
\vcenteredhbox{\resizebox{0.4\textwidth}{!}{\includegraphics{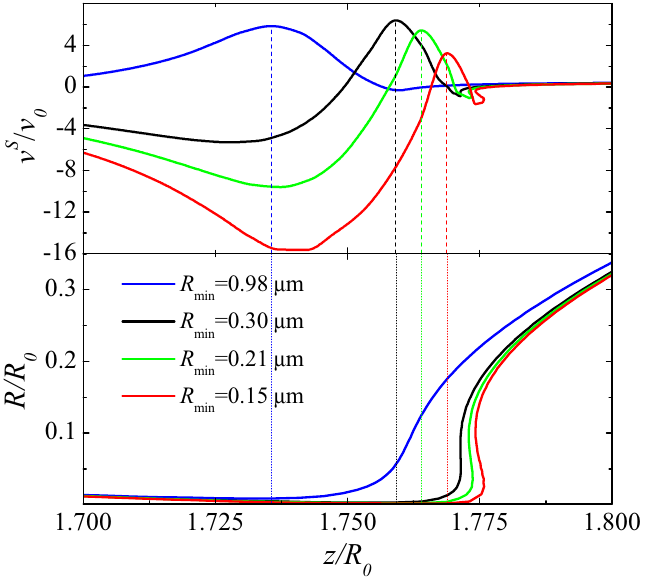}}}
\caption{Surface velocity $v^S(z)$ (a) and free surface radius $R(z)$ (b) for DIW+SDS 0.8cmc with $\{\mu_1^{S*}=0$, $\mu_2^{S*}=3.5\times 10^{-9}$ Pa\,s\,m$\}$. The dashed vertical lines indicate the position of the free surface neck.}
\label{vs}
\end{figure}

Figure \ref{vs} shows the velocity ${\bf v}^S=v^S {\bf t}$ along the free surface as the droplet approaches its breakup for the case $\{\mu_1^{S*}=0$, $\mu_2^{S*}=3.5\times 10^{-9}$ Pa\,s\,m$\}$. As can be observed, the maximum of $v_s(z)$ exhibits a non-monotonic behavior with respect to the time to the pinching, and is located at the free surface neck. The difference between the maximum and minimum values of $v_s(z)$ increases with time, and so does the average dilatational stress in the pinching region. The overturning of the free surface is observed for $R_{\textin{min}}\lesssim 0.3$ $\mu$m. For this reason, $v_s(z)$ becomes a multivalued function on the right side of the free surface neck.

We now study how the scaling of the minimum radius depends on the surfactant viscosities. In general, we have $R_{\textin{min}}=f(\tau,\mu_{1,2}^S)$. Assume that we can write this equation in the form $R_{\textin{min}}=R_s H(\tau/\tau_s)$, where $R_s$ and $\tau_s$ are the length and time scales associated with the surface viscosities, respectively. We suppose that these scales depend on the viscosities as 
\begin{equation}
\label{scaling}
R_s=A (\mu_{1,2}^{S*})^{\alpha}, \quad\tau_s=B (\mu_{1,2}^{S*})^{\beta}.    
\end{equation}
The cross-over function $H(\xi)$ behaves as $H(\xi)\sim\xi^{2/3}$ for $\xi\gg 1$ (inviscid limit) and $H(\xi)\sim\xi^{\gamma}$ for $\xi\ll 1$ (viscous regime), with a crossover at $\xi\sim 1$. Therefore, $R_{\textin{min}}=A B^{-2/3} (\mu_{1,2}^S)^{\alpha-2\beta/3}\tau^{2/3}$ in the inviscid limit. Assuming that $R_{\textin{min}}\sim \tau^{2/3}$ in that limit, we conclude that $\alpha=2\beta/3$.

The value of the exponents can be guessed from the balance of forces. Both Marangoni and surface viscous stresses delay the free surface pinch-off (Figs.\ \ref{W08} and \ref{W2}) acting against the driving capillary force. For sufficiently small values of $R_{\textin{min}}$, the effect of surface viscous stresses become comparable and even larger than that caused by Marangoni stress (Figs.\ \ref{distributionshear} and \ref{distributiondilatational}). The value of $R_{\textin{min}}$ below which this occurs decreases as the surface viscosities decrease. For instance, Marangoni and surface viscous stresses produce similar effects for $R_{\textin{min}}\lesssim 2$ $\mu$m and $R_{\textin{min}}\lesssim 0.15$ $\mu$m in the cases $\{\mu_1^{S*}=0$, $\mu_2^{S*}=10^{-7}$ Pa\,s\,m$\}$ and  $\{\mu_1^{S*}=0$, $\mu_2^{S*}=3.5\times 10^{-9}$ Pa\,s\,m$\}$, respectively. Therefore, we expect surface viscous stresses to be commensurate with the driving capillary pressure in the pinch-off region for those intervals of $R_{\textin{min}}$. In fact, the interfacial Ohnesorge numbers $\text{Oh}_{1,2}^{S*}$ defined in terms of $R_{\textin{min}}$ take values at least of order of unity in those intervals. 

The balance between the capillary pressure and the surface viscous stresses in Eq.\ (\ref{NormalStress1}) yields 
$\sigma_0/R_s\sim \mu_{1,2}^{S*}/(R_s\tau_s)$, where we have taken into account that the variation of surface velocity scales as $(R_s/\tau_s)/R_s$ due to the continuity equation. The above balance allows us to conclude that $\beta=1$, and therefore $\alpha=2/3$. According to our analysis, 
\begin{equation}
\frac{R_{\textin{min}}}{(\mu_{1,2}^{S*})^{2/3}}\sim \left(\frac{\tau}{\mu_{1,2}^{S*}}\right)^{\gamma}
\end{equation}
in the viscous regime. 

In the 1D (slenderness) approximation \citep{E97}, the axial forces per unit volume due to the shear and dilatational surface viscosities are $(9\mu_1^S R w_z)_z/2R^{2}$ and $(\mu_2^S R w_z)_z/2R^{2}$ \citep{MS18}, respectively, where $w$ is the $z$-component of the velocity and the subscript $z$ indicates the derivative with respect to the coordinate $z$. As can be seen, the terms corresponding to the shear and dilatational viscosities differ only by a factor 9. Therefore, the asymptotic behavior of $R_{\textin{min}}(\tau)$ for $\{\mu_1^{S*}=a$, $\mu_2^{S*}=0\}$ ($a$ is an arbitrary constant) is expected to be the same as that for $\{\mu_1^{S*}=0$, $\mu_2^{S*}=9a\}$. As will be seen below, this allows us to group the simulation results for $\mu_1^{S*}\neq 0$ and $\mu_2^{S*}\neq 0$. 

\begin{figure}[h]
\vcenteredhbox{\resizebox{0.425\textwidth}{!}{\includegraphics{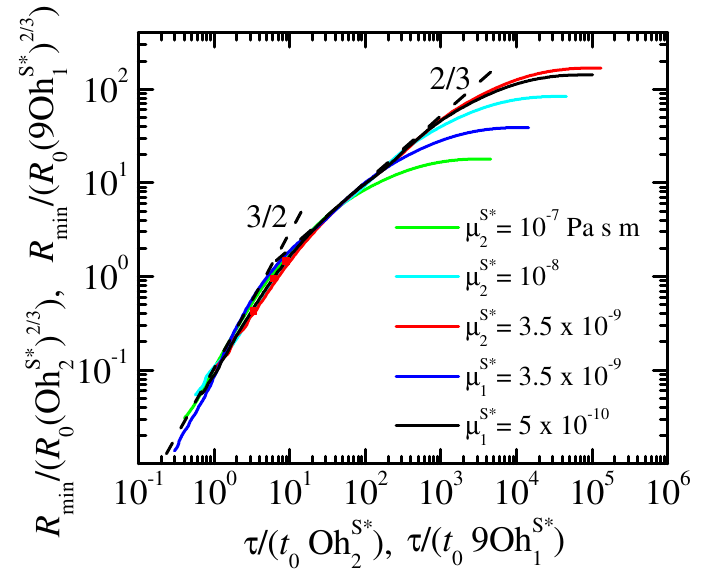}}}
\caption{Dimensionless minimum radius $R_{\textin{min}}/R_0$ as a function of the dimensionless time to the breakup, $\tau/t_0$, for the breakup of a pendant drop of DIW+SDS 0.8cmc. The labels indicate the values of the nonzero shear/dilatational viscosity in each case.}
\label{ssr}
\end{figure}

Using the equivalence $9\mu_1^S\leftrightarrow \mu_2^S$, we find the values of the exponents $\beta$ and $\gamma$ leading to the collapse of all the numerical data for $R_{\textin{min}}\to 0$. Following the optimization method described by \citet{MG20}, the best collapse is obtained for $\beta=1.1$ and $\gamma=1.4$. Figure \ref{ssr} shows the results scaled with the exponents $\beta=1$ and $\alpha=2/3$ calculated in the previous analysis. As explained above, we have grouped the results for nonzero shear and dilatational viscosities using the factor 9 suggested by the 1D model. The simulations show the transition from the inertio-capillary regime $R_{\textin{min}}\sim \tau^{2/3}$ to the asymptotic behavior given by power law $\gamma=3/2$.

\begin{figure}[h]
\vcenteredhbox{\resizebox{0.35\textwidth}{!}{\includegraphics{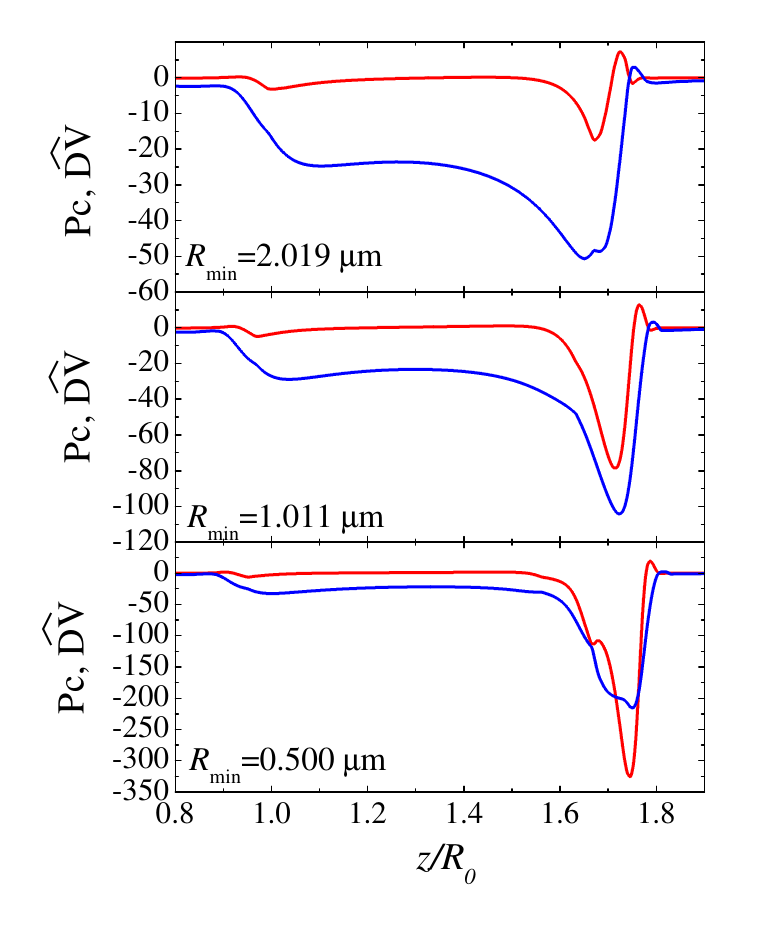}}}\vcenteredhbox{\resizebox{0.36\textwidth}{!}{\includegraphics{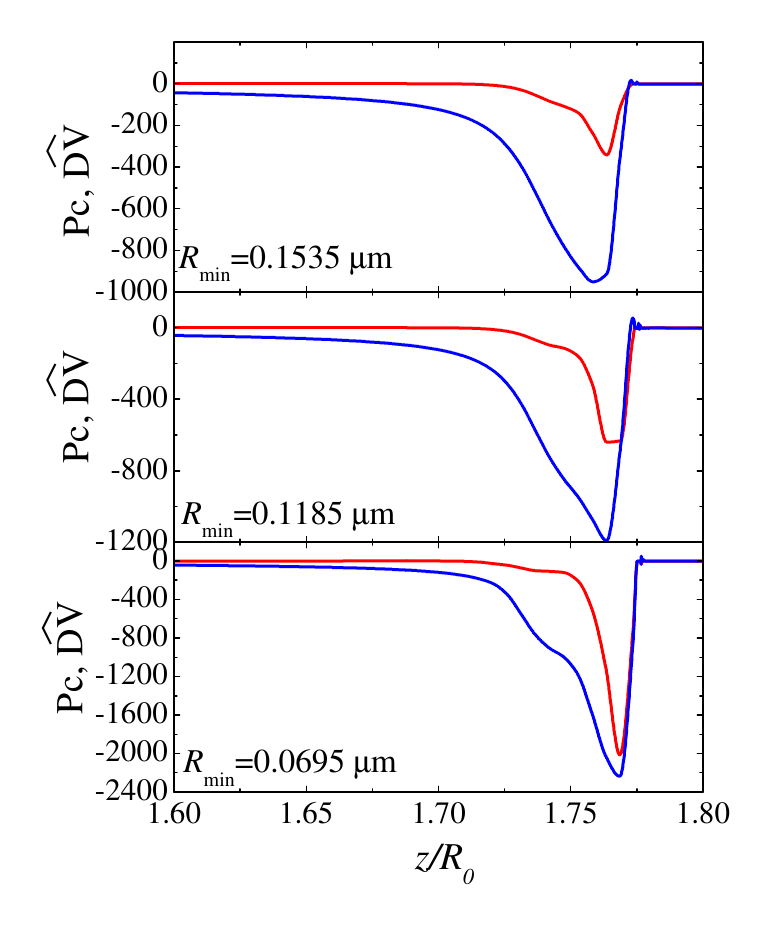}}}
\caption{Axial distribution of the capillary stress $\text{Pc}=\hat{\sigma}\kappa$ (blue lines) and normal dilatational viscous stress $\widehat{\text{DV}}=\text{Oh}_2^S(\boldsymbol{\nabla}^S\cdot\mathbf{v}^S) \kappa$ (red lines) for DIW+SDS 0.8cmc and three instants as indicated by the value of $R_{\textin{min}}$. The left-hand and right-hand graphs correspond to $\{\mu_1^{S*}=0$, $\mu_2^{S*}=10^{-7}$ Pa\,s\,m$\}$ and  $\{\mu_1^{S*}=0$, $\mu_2^{S*}=3.5\times 10^{-9}$ Pa\,s\,m$\}$, respectively.}
\label{stresses}
\end{figure}

The axial distributions of the capillary pressure and the dilatational viscous stress are shown in Figs.\ \ref{stresses} for the cases $\{\mu_1^{S*}=0$, $\mu_2^{S*}=10^{-7}$ Pa\,s\,m$\}$ and  $\{\mu_1^{S*}=0$, $\mu_2^{S*}=3.5\times 10^{-9}$ Pa\,s\,m$\}$. As can be observed, the dilatational viscous stress becomes comparable with the driving capillary pressure for $R_{\textin{min}}\lesssim 2$ $\mu$m and $R_{\textin{min}}\lesssim 0.15$ $\mu$m in the cases $\mu_2^{S*}=10^{-7}$ Pa\,s\,m and $\mu_2^{S*}=3.5\times 10^{-9}$ Pa\,s\,m, respectively. This explains the good agreement between the numerical simulations and the scaling proposed above for the minimum radius.

\section{Conclusions}
\label{sec5}

We studied both numerically and experimentally the breakup of a pendant water droplet loaded with SDS. We measured a delay of the droplet breakup with respect to that predicted when only solutocapillarity and Marangoni stress ate accounted for. This delay is attributed to the role played by surface viscosities. When Marangoni and surface viscosity stresses are accounted for, then surface convection does not sweep away the surfactant from the thread neck, at least in the time interval analyzed. The results show that surface viscous stresses have little influence on both the surfactant distribution along the free surface and the free surface position. Therefore, the size of the satellite droplet and the amount of surfactant accumulated in it are hardly affected by the surface viscosities. These results differ from those obtained for a much more viscous surfactant \citep{PMHVV17}. As the free surface approaches its breakup, an inertio-capillary regime gives rise to that in which surface viscous stresses become commensurate with the driving capillary pressure. We have proposed a scaling law to account for the effect of surface viscosities on $R_{\textin{min}}(\tau)$ in this last regime.

The pinching of an interface is a singular phenomenon that allows us to test theoretical models under extreme conditions. The vanishing spatiotemporal scales reached by the system as the interface approaches its breakup unveil physical effects hidden in phenomena occurring on much larger scales. This work is an example of this. Surface viscous stresses become relevant in the vicinity of the pinching region long before thermal fluctuations become significant \citep{ML00,E02}, even for practically inviscid surfactants, such as SDS. In this sense, the surfactant-laden pendant droplet can be seen as a very sensitive surfactometer to determine the values of the surface viscosities, which constitutes a difficult problem \citep{ELS16}. A series of experiments for different surfactant concentrations and needle radii may lead to accurate measurements of $\mu_1^{S}(\Gamma)$ and $\mu_2^{S}(\Gamma)$ characterizing the behavior of low-viscosity surfactants.

\vspace{1cm}
Partial support from the Ministerio de Econom\'{\i}a y Competitividad and by Junta de Extremadura (Spain) through Grant Nos. DPI2016-78887 and GR18175is gratefully acknowledged.



%

\end{document}